\documentclass[graybox]{svmult}

\usepackage{amsfonts}
\usepackage{type1cm}        % activate if the above 3 fonts are
\usepackage{makeidx}         % allows index generation
\usepackage{graphicx}        % standard LaTeX graphics tool
\usepackage{multicol}        % used for the two-column index
\usepackage[bottom]{footmisc}% places footnotes at page bottom

\usepackage{newtxtext}       % 
\usepackage{newtxmath}       % selects Times Roman as basic font
\usepackage{amsmath}

\usepackage{listings}
\usepackage{multirow}
\usepackage{soul}
\usepackage{xspace}

\usepackage{tikz}
\newcommand*\circled[1]{\tikz[baseline=(char.base)]{
            \node[shape=circle,draw,inner sep=1pt] (char) {#1};}}

\makeindex             % used for the subject index
\definecolor{gr}{rgb}{0.0, 0.5, 0.0}
\definecolor{b}{rgb}{0.0, 0.0, 1.0}
\definecolor{lb}{rgb}{0.0, 0.0, 0.5}
\definecolor{purple}{rgb}{0.55, 0.0, 0.55}
\definecolor{eclipseBlue}{RGB}{42,0.0,255}
\definecolor{eclipseGreen}{RGB}{63,127,95}
\definecolor{eclipsePurple}{RGB}{127,0,85}
\definecolor{lbcolor}{rgb}{0.9,0.9,0.9}  
\definecolor{lightblue}{rgb}{0.68, 0.85, 0.9}

\lstdefinelanguage{SPARQL}{
  showstringspaces=false,
  comment=[l]{//},
  morecomment=[s]{/*}{*/},
  commentstyle=\color{gr}\ttfamily,
  morecomment=[l]{\#},       % comments
  morecomment=[n][\color{lb}]{<http}{>}, % uris
  morestring=[b][\color{green}]{\"},  % strings
  classoffset=1,
  morekeywords={?O, sh, sosa, owl, xsd, purl, ssr}, keywordstyle=\color{b},
  classoffset=2,
  morekeywords={
    SELECT,CONSTRUCT,DESCRIBE,ASK,WHERE,FROM,NAMED,PREFIX,BASE,OPTIONAL,
    FILTER,GRAPH,LIMIT,OFFSET,SERVICE,UNION,EXISTS,NOT,BINDINGS,MINUS,
    STREAM, NAF, WINDOW
  },
  keywordstyle=\color{gr}\textbf,
  classoffset=3,
  morekeywords={
   det, det1, det2, det3, det4, det5, det8, det11, car, hasScore, resultTime, 
   isSampleOf, usedProcedure, Kalman, isResultOf, cam, madeBySensor, Sampling, Observation, hasSimpleResult, isDetectionOf,
   image2, Image2D, obs2, Yolo, KalmanFilter,
   b1,b2,b3,b4,b5,b6,b7,b8,b9,b10,b11,
   trk1, trk2, trk3, trk,trk23,trk5,
   rule_w_1,rule_w_2, rule_w_3, rule, leaves, inFOV, rule_w_4, vMatch, ends,
   NodeShape, FoV, CQELSRule, prefixes, construct, rule_w_5, score,
   enters, trklet
  },
  keywordstyle=\color{purple},
  classoffset=4,
  otherkeywords={},
  morekeywords={
    window, sec, iou, @, &&
  },
  keywordstyle=\color{red},
  classoffset=5,
  morekeywords={a}, keywordstyle=\color{lb}\textbf,
}

\lstdefinelanguage{asp}{
  keywords={not, at,CONSTRUCT,WHERE,NAF,STREAM,window,FILTER,from,in, sosa},
  otherkeywords = {:-,@,[, ]},
  keywordstyle={\color{blue}},
  ndkeywords={pred, leaves, det, iSO, trajectory, enters,trklet, ends, starts, reid, inFoV, trk, iou,  score, vMatch,hasConfScore,isSampleOf, initiates, terminates},
  ndkeywordstyle=\color{lb},
  sensitive=false,
  comment=[l]{//},
  morecomment=[s]{/*}{*/},
  commentstyle=\color{gr}\ttfamily,
  stringstyle=\color{red}\ttfamily,
  morestring=[b]',
  classoffset=2,
  keywordstyle=\color{red},
  morekeywords={car},
}
\newcommand{\system}{\mbox{\sffamily{CQELS~2.0}}\xspace}%      
\newcommand{\name}[1]{\mbox{\ttfamily{#1}}\xspace}%  

\begin{document}

\title*{CQELS 2.0: Towards A Unified Framework for Semantic Stream Fusion }
\author{Anh Le-Tuan, Manh Nguyen-Duc, Chien-Quang Le, Trung-Kien Tran,\newline Manfred Hauswirth, Thomas Eiter and Danh Le-Phuoc}
% Use 
\authorrunning{Anh Le-Tuan et al. } 
\institute{Anh Le-Tuan \and Manh Nguyen-Duc \at Open Distirbuted Systems, Technical University of Berlin, Germany
\and Chien-Quang Le \at University of Science, Hue, Vietnam
\and Trung-Kien Tran \at Bosch Center for Artificial Intelligence, Renningen, Germany  
\and Thomas Eiter \at  Vienna University of Technology
\and Manfred Hauswirth \and Danh Le-Phuoc \at Open Distirbuted Systems, Technical University of Berlin\newline Fraunhofer Institute for Open Communication Systems, Berlin, Germany}

\maketitle

\abstract
{We present \system, the second version of \textbf{C}ontinuous \textbf{Q}uery \textbf{E}valuation over \textbf{L}inked \textbf{S}treams. \system is a platform-agnostic federated execution framework towards semantic stream fusion. In this version, we introduce a novel neural-symbolic stream reasoning component that enables specifying deep neural network (DNN) based data fusion pipelines via logic rules with learnable probabilistic degrees as weights. As a platform-agnostic framework, \system can be implemented for devices with different hardware architectures (from embedded devices to cloud infrastructures). Moreover, this version also includes an adaptive federator that allows CQELS instances on different nodes in a network to coordinate their resources to distribute processing pipelines by delegating partial workloads to their peers via subscribing continuous queries.}

%\section{A Call for An Overhaul of CQELS Framework}
\section{Introduction and Motivation}

% CQELS started about in 2011~\cite{Danh:2011} as one of the first RDF Stream processing engines to solve the data integration problem for sensor data, Internet of Things, and the Web of data. The key selling point of this data integration approach is the semantic interoperabiliy enabled by RDF data models. The framework has gone through different development phases which were driven by use cases from industry, EU,  reserarch projects funded by EU, Ireland and Germany. For instance, one of its implementations is licensed to commercial IoT Gateways even the general framework is open sourced.

CQELS was proposed as one of the first RDF stream processing engines to solve the data integration problem for sensor data, Internet of Things, and the Web of data~\cite{Danh:2011}. The key feature %selling point 
of this data integration approach is the semantic interoperabiliy enabled by RDF data models.
% The framework has gone through different development phases which were driven by use cases from industry, EU,  research projects funded by EU, Ireland and Germany. For instance, one of its implementations is licensed to commercial IoT Gateways even the general framework is open sourced.
The framework has been used in various industry and research applications both as open source and commercial software.% e.g. IoT Gateways\kien{please add citation}.

% After ten years, we are integrating an increasing number of stream data types  such as video streams, LiDARs, and we are enabling more hardwares such as ARM and mobiles in ~\cite{} . Moreover, we supported for data fusion operations in our engines~\cite{}. The recent advances in \textit{machine learning} (ML) with \textit{deep neural network} (DNN) enable extracting a rich set of (visual) features from multimodal stream data with high accuracy. For example, object detection DNNs~\cite{Joseph:2015, Shaoqing:2015} return a set of bounding boxes and object classes given an image or frame of video. Thus,  CQELS 2.0, the second version of CQELS~\cite{Danh:2011},  is designed as a DNN-based data stream fusion framework. Along with accepting different data sources, the CQELS-QL query language has evolved from supporting event query patterns~\cite{Minh:2015,Daniele:2016} to probalistic reasoning in~\cite{Manh:2021,Danh:2021}.

The recent advances in \textit{machine learning} (ML) with \textit{deep neural network} (DNN) leads to the enormous amount of data in various formats, and in many cases from multimodal stream data with high accuracy. Therefore, the existing framework is no longer suitable for the new application scenarios and requires significant extensions. Towards this goal, in \system, we integrate a number of new stream data types such as video streams, LiDARs, and support more hardwares such as ARM and mobiles~\cite{Anh:2018,Manh:2019}. Moreover, we provide data fusion operations in our engines~\cite{Danh:2021,Manh:2021}. For example, the operation for object detection DNNs~\cite{Joseph:2015, Shaoqing:2015} returns a set of bounding boxes and object classes given an image or video frames. Thus,  \system is designed as a DNN-based data stream fusion framework. Along with the new feature for different data sources, the CQELS-QL query language has evolved from supporting event query patterns~\cite{Minh:2015,Daniele:2016} to probabilistic reasoning in~\cite{Danh:2021,Manh:2021}.

% This leads us to the quest to take the an important architecture and design overhaul of CQELS framework to meet the implementation demand and research in coming years. The new design is taking inputs and requirements from use cases and research problems from many research projects (e.g. DFG COSMO and BMBF BIFOLD) and industry partners. For example, application scenarios around edge intelligence and industry 4.0 from DellEMC, Siemens and Bossh movivated our effort in building autonomous processing kernels powered by CQELS in~\cite{Manh:2019,Manh:2021}.  Such systems will deal with city-scale camera deployments which have become ubiquitous in smart cities and traffic monitoring with a steadily increasing in size and reach of their deployments. For instance, the British Security Industry Association estimates that there are between 4 to-5.9 million CCTV cameras in UK~\cite{cctv}. To this end, the new design of CQELS framework has deal with not only the new level of complexity of data sources, processing operations and deployment settings but also with much bigger processing scale in terms of throughput and volume.

To implement such complex features and also to support for future demand in research, we bring the existing CQELS framework to the next level with new architecture and design. The new design is driven by requirements from use cases and research problems from various research projects (e.g. DFG COSMO\footnote{https://gepris.dfg.de/gepris/projekt/453130567?language=en} and BMBF BIFOLD\footnote{https://bifold.berlin/}) and industry partners. For example, application scenarios in edge intelligence~\cite{Zhi:2019,Shuiguang:2020} and industry 4.0 from DellEMC, Siemens and Bossh motivate us to build \emph{autonomous processing kernels} powered by CQELS~\cite{Manh:2019,Manh:2021}. Such systems deal with city-scale camera deployments which have become ubiquitous in smart cities and traffic monitoring with a continuously increasing in size and utilization of their deployments. For instance, the British Security Industry Association estimates that there are between 4 to 5.9 millions CCTV cameras in the UK~\cite{cctv}. To this end, the new design of CQELS framework has to deal with not only the new level of complexity of data sources, processing operations and deployment settings, but also with significantly larger processing scale in terms of throughput and volume.

\section{CQELS 2.0 Framework for Semantic Stream Fusion}

\begin{figure}[ht!]
    \centering
    \includegraphics[width=1\textwidth]{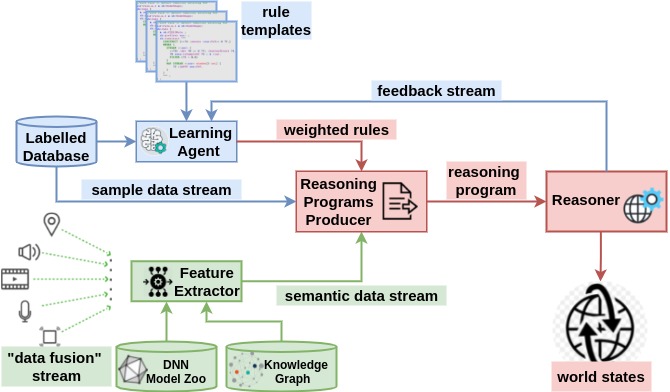}
    \caption{The overview of conceptual design of CQELS 2.0}
    \label{fig:concept}
\end{figure}

Semantic stream reasoning (SSR)~\cite{Danh:2021} enables the data fusion of multimodal stream data such as camera and LiDARs via declarative rules. Such rules can be written in SPARQL-based or Answer Set Programming (ASP) syntax. And the stream data flow between data fusion operations can be represented as standardized data formats, e.g, RDF Star. SSR generalizes the data model and processing operations of the previous CQELS engines, hence,  CQELS 2.0 framework uses SSR formalisation and abstractions to facilitate the architecture design and implementation. Figure~\ref{fig:concept} illustrates the conceptual design of CQELS 2.0 framework that comprises several components. The~\name{Feature Extractor} component extracts the features of interest from the incoming information streams and maps them to the symbolic representation. The feature data is described with a~\emph{neuro-symbolic stream model}~\cite{Danh:2021} and its semantic is enriched via linked %being linked to 
knowledge graphs.

The~\name{Reasoning Programs Producer} component takes responsibility to generate \emph{semantic stream reasoning programs}~\cite{Danh:2021} which specify the fusion pipeline and the decision logic to choose the most likely state of the world at each evaluation. The reasoning program is evaluated by the \name{Reasoner} component, which employs an ASP solver. There are two types of reasoning rules: \emph{hard rules} and \emph{soft rules}. The hard rule is used for background knowledge given by (non-monotonic) common-sense and domain knowledge that is regarded as "always true". The soft rules expresses association hypotheses with weights corresponding to probability degrees of these rules. The weights of the rules are determined by the \name{Learning Agent} component in the starting phase.

The workflow of the framework is as follows. For the setup, the symbolic training samples are constructed from the labeled data and are fed to the \name{Learning Agent} component. The \name{Learning Agent} component computes a vector of weights for the soft rules which are stored as rule templates and passes them to the \name{Reasoning Programs Producer} component. For each training iteration, a \emph{feedback stream} returns the reasoning results back to \name{Learning Agent} component. 
The \name{Learning Agent} component adjusts the weights of the soft rules until the answer sets (returned as feedback streams) describe the most likely ground truth. 
%\kien{this sentence is not so clear to me, especially this part "until the outputs a stream of answer sets describing..."}.

\section{Provisional Features}

In this section, we present the details of a provisional feature of \system that can be used to solve the \emph{multiple object tracking} (MOT) problem~\cite{Bernardin:2008}.
In computer vision, MOT algorithms are normally programmed in C/C++ or Python. 
With \system, we will show that the widely used tracking-by-detection approach~\cite{Ciaparrone:2019} for the MOT problem can be emulated in a declarative fashion with rules and queries.

The key operations in tracking-by-detection approaches (e.g., SORT~\cite{Bewley:2016}, DEEPSORT~\cite{Wojke:2017}) are as following: 1) detection of objects (using DDN-based detector), 2) propagating object states (e.g., location and velocity) into future frames, 3) associating current detection with existing% objects, and 4) managing the lifespan of the tracked objects. 

\begin{figure}[ht!]
    \centering
    \includegraphics[width=1.\textwidth]{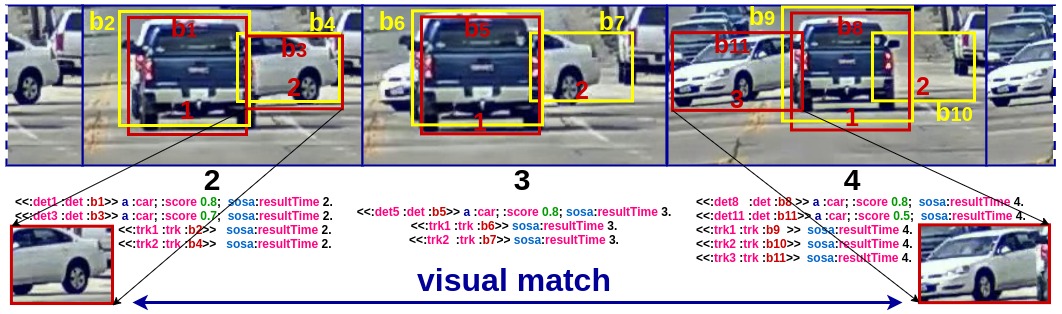}
    \caption{A Semantic Visual Stream Snapshot. The red boxes are detected bounding boxes and the yellow boxes are tracked bounding boxes.}
    \label{fig:example}
\end{figure}

Figure~\ref{fig:example} illustrates the SORT algorithm~\cite{Bewley:2016} which is a simple object tracking algorithm based on DNN detectors such as SSD~\cite{Liu:2016} or YOLO~\cite{Redmon:2017}. 
To associate resultant detections with existing targets, SORT uses a Kalman filter~\cite{Bishop:2001} to predict the new locations of targets in the current frame. 
At time point 2, the red boxes $b_1$ and $b_3$ are newly detected, and the yellow boxes $b_2$ and $b_4$ are predicted by a Kalman filter based on the tracked boxes from the previous frame. 
Then, the SORT algorithm computes an associative cost matrix between detections and targets based on the intersection-over-union (IOU) distance between each detection and all predicted bounding boxes from the existing tracklets.
In case some detection is associated with a target, the detected bounding box is used to update the target state via the Kalman filter. 
As in frame 2, the tracklets $trk_1$ and $trk_2$ are set to the two new bounding boxex $b_1$ and $b_2$ which are associated with predicted boxes $b_3$ and $b_4$ respectively. Otherwise, the target state is simply predicted without correction using the linear velocity model. For example, at time point 3, the detector misses detecting the white car due to an occlusion, however, the tracklet 2 is till assigned to box $b_7$ which contains part of the white car.

\begin{figure}[ht!]
    \centering
    \includegraphics[width=1\textwidth]{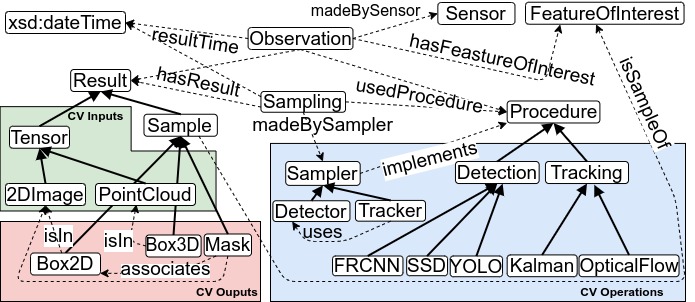}
    \caption{Data abstraction of Neural-Symbolic stream with Semantic Sensor Network Ontology}
    \label{fig:abstract}
\end{figure}

With the Feature Extractor component (see. Figure~\ref{fig:concept}), \system extracts the detection boxes from input video frames and represent them as RDF facts. The data abstraction is modelled using the neural symbolic stream modelling practice (presented in~\cite{Danh:2021}), which extends the standardised Semantic Sensor Network Ontology (SSN)~\cite{Armin:2019} to model video streams and intermediate processing states. The extension includes various vocabularies to specify the semantics of camera sensors, video frames, and tensors as shown in Figure~\ref{fig:abstract}. 

With the above data abstraction, \system consumes the video data stream that is observed by a camera (as a ${Sensor}$) as \emph{a stream of observations}. Each observation is a video frame. These video frames are represented as instances of the class ${Image2D}$ that inherits from the generic ${Observation}$ class of SSN. The detection of a video frame or a tracklet are represented as SSN $Sampling$ which are processed by a DNN model or a CV algorithm represented as a \emph{Procedure}.

\begin{lstlisting}[
caption={Semantic Stream Serialization with RDF$^*$}, 
label={rdf:frame2},
language=SPARQL,
numbers=left,
captionpos=b,
basicstyle=\footnotesize\ttfamily,  
backgroundcolor=\color{lbcolor},]
//time point/frame 2
<<:image2 a :Image2D>> a sosa:Observation; 
sosa:madeBySensor :cam1; sosa:resultTime 2.

<<:det1 :det :b1>> a :Detection; sosa:resultTime 2;
sosa:hasSimpleResult 'car'; :score '0.8';
:isDetectionOf :image2; sosa:usedProcedure :Yolo.

<<:det1 :det :b3>> a :Detection; sosa:resultTime 2;
sosa:hasSimpleResult 'car'; :score '0.7';
:isDetectionOf :image2; sosa:usedProcedure :Yolo.

<<:trk1 :trk :b2>> a :Tracklet; sosa:resultTime 2;
sosa:usedProcedure :KalmanFilter.

<<:trk2 :trk :b4>> a :Tracklet; sosa:resultTime 2;
sosa:usedProcedure :KalmanFilter.
\end{lstlisting}

For example, Listing~\ref{rdf:frame2} illustrates the the symbolically representation in RDF* format\footnote{\url{https://w3c.github.io/rdf-star/cg-spec/editors$\_$draft.html}} of the frame 2 in the stream snapshot in Figure~\ref{fig:example} Line 5 denotes that the detection model generates an output consisting of a bounding box $b_1$, object type $car$ and confidence score $0.8$. Line 13 presents that the box $b_2$ is predicted by a Kalman filter and is tracked by tracklet 1.

To emulate these MOT algorithms, \system represents these association hypotheses by hard rules and soft rules that are translated into an optimization problem solved by an ASP Solver. To formalise the reasoning process on semantic representations of stream data, we use a temporal semantic model that allows us to reason about the properties and features of objects. The model accounts for the laws of the physical world and commonsense, which allow the system to handle incomplete information (e.g., if we do not see objects appearing in observations, or camera reads are missing).

Using the above RDF representations, we allow RDF\&SPARQL developers to write soft rules with SHACL rule language\footnote{SHACL Advanced Features} along with the extension of CQEL-QL. 
The extension we made here is replacing SPARQL CONSTRUCT with the  corresponding CONSTRUCT of CQELS-QL which extends SPARQL with the window operators over RDF Stream. 
Moreover,  extending SHACL for  expressing ASP-like rules is aligned  with the recent proposal for assigning SHACL to negation stable semantics~\cite{Medina:2020}.
Note that we extend CQESL-QL syntax in~\cite{Danh:2011} with the keyword "NAF" to express the default negation of ASP.
For example, the soft rule 1 in Listing~\ref{soft-rule1} is used to trigger the event a car enters the "Field of View" of a camera and the car starts being tracked. 

\begin{lstlisting}[
caption={Soft rule 1 - detect vehicles entering Field of View}, 
label={soft-rule1},
captionpos=b,
language=SPARQL,  
numbers=left,
basicstyle=\footnotesize\ttfamily,  
backgroundcolor=\color{lbcolor},
escapeinside=||,]
ssr:rule_w_1 a sh:NodeShape;
sh:rule [
  a sh:CQELSRule ;
  sh:prefixes ssr: ;
  sh:construct |"""|
    CONSTRUCT {<<?O :enters <ssr:FoV>> @ ?T.}
    WHERE {
      STREAM <:ssr> {
        <<?Dt :det ?B >> @ ?T; :score ?S.
        ?B sosa:isSampleOf ?O ; a :car.
        FILTER (?S > 0.8)
      }
      NAF STREAM <:ssr> window[5 sec] {
          ?O :inFOV ssr:FoV.
      }
    }
    |"""| ;
] ;
\end{lstlisting}

To associate a detected bounding box $B$ with an object $O$, we use soft rules that assert the triple <<B sosa:isSampleOf O>> based on {\em explained spatial, temporal, and visual appearance evidences}. Such rules can be used to represent {\em hypotheses}\/ on temporal relations among detected objects in video frames following a tracking trajectory. When the object's movement is consistent with the constant velocity model, e.g., the Kalman filter used in SORT~\cite{Bewley:2016}, and there is a detection associated with its trajectory, the fact <<B sosa:isSampleOf O>> is generated by rule~\ref{soft-rule2}.
Here, $iou(B_1,B_2)$ states the IOU (intersection over union) condition of the bounding boxes $B_1$ and $B_2$ satisfy.

\begin{lstlisting}[
caption={Soft rule 2 - emulation of SORT algorithm~\cite{Bewley:2016}.}, 
label={soft-rule2},
captionpos=b,
language=SPARQL,  
numbers=left,
basicstyle=\footnotesize\ttfamily,  
backgroundcolor=\color{lbcolor},
escapeinside=||,]
ssr:rule_w_2 a sh:NodeShape ;
sh:rule [
  a sh:CQELSRule ;
  sh:prefixes ssr: ;
  sh:construct |"""|
  CONSTRUCT { ?B1 sosa:isSampleOf ?O. }
  WHERE{
    STREAM <:ssr>{
      <<?Dt :det ?B2 >> @ ?T; :score ?S.
      <<?Trk :trk ?B1 >> @ ?T.
      ?Trk :trklet ?O.
      FILTER (?S>0.8 && iou (?B1,?B2) > 0.8)
    }
  }
  |"""| ;
] ;
\end{lstlisting}

\begin{lstlisting}[
caption={Soft rule 3 - emulation of DEEPSORT algorithm~\cite{Wojke:2017}.}, 
label={soft-rule3},
captionpos=b,
language=SPARQL,  
numbers=left,
basicstyle=\footnotesize\ttfamily,  
backgroundcolor=\color{lbcolor},
escapeinside=||,]
ssr:rule_w_3 a sh:NodeShape ;
sh:rule [
  a sh:CQELSRule ;
  sh:prefixes ssr: ;
  sh:construct |"""|
  CONSTRUCT { ?B1 sosa:isSampleOf ?O. }
  WHERE{
    STREAM <:ssr> @?Te window[5 sec] {
      ?Trk2 :trk ?B2
    }
    STREAM <:ssr>{
      <<?Trk1 :trk ?B1 >> @ ?T.
      <<?B1 :vMatch ?B2 >> :score ?S.
      ?B2 sosa:isSampleOf ?O.
      ?Trk2 :ends ?Te.
      FILTER {?T<?Te+3 && ?S>0.8 }
    }
  }
  |"""| ;
] ;	
\end{lstlisting}

Furthermore, we can also emulate DeepSORT~\cite{Wojke:2017} via the soft rule~\ref{soft-rule3} that can search for supporting evidences to link a newly detected bounding box from an occluded tracklet using visual appearance associations, e.g. frames 2 and 4 of Figure~\ref{fig:example}. 
For this, we search for pairs of bounding boxes from recently occluded tracklets w.r.t. visual appearance. As the search space of possible matches is large, we limit it by filtering the candidates based on their temporal and spatial properties. To this end, we use rules with windows to reason about disconnected tracklets that have bounding boxes visually matched within a window of time points that are aligned with DeepSORT's gallery of associated appearance descriptors for each tracklet. Based on this gallery of previously tracked boxes, the appearance-based discriminative metrics are  computed to recover the identities after long-term occlusion, where the motion is less discriminative. Hence, to connect a newly detected bounding box $B_1$ that has a visual appearance match with another bounding box $B_2$ (represented  by the fact <<$B_1$ :vMatch $B_2.$>> in line 13) of a discontinued tracklet $T_2$ (represented by   <<$Trk_2$ :ends $T_e$>> in line 15) that ended 3 time points before.
\section{System Architecture Overview}
In this section, we present a specific architecture of \system for the conceptual design in Figure~\ref{fig:concept}. 
CQELS framework provides a platform-independent infrastructure to implement RDF-Stream Processing (RSP) engines for computing continuous
queries expressed in CQELS-QL. The first version of CQELS accepts RDF streams as input and returns RDF streams or relational streams in the SPARQL format as output. \system allows creating RDF streams by annotating extracted features from other data streams. The output RDF streams can be fed into any RSP engine, and the relational stream can be used by other relational stream processing systems.

\begin{figure}[ht!]
    \centering
    \includegraphics[width=1\textwidth]{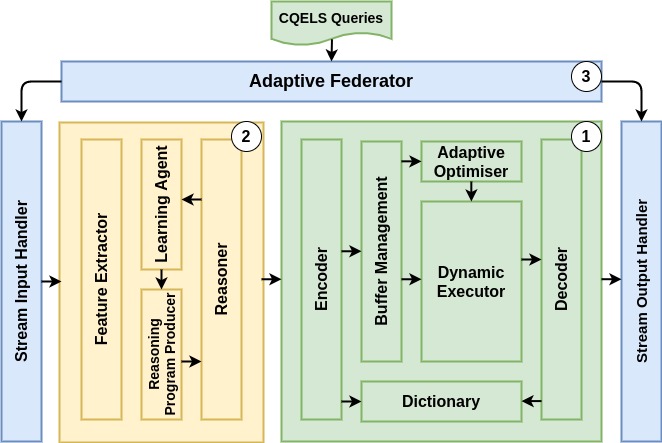}
    \caption{System architecture of \system.}
    \label{fig:cqels20}
\end{figure}

Figure~\ref{fig:cqels20} illustrates the overview architecture of \system. In general, \system consists of three subsystems.
Subsystem \circled{1}, the RDF stream processing processor,  extends the stream processing primitives of its previous version~\cite{Danh:2011} to accelerate the grounding phase of stream reasoning. For example, multiway joins are used to accelerate the incremental ground techniques~\cite{Danh:2018}.
The second subsystem~\circled{2} is the semantic stream reasoning component as presented in Section 2. Finally, subsystem~\circled{3} is a stream adaptive federator, which is described as follows.

Thanks to the platform-agnostic design of its execution framework~\cite{Danh:2015}, the core components are abstract enough to be seamlessly integrated with different RDF libraries in order to port the resulting system to different hardware platforms. For scalability, CQELS employs Storm\footnote{Storm. https://storm.apache.org/} and HBase\footnote{Hbase. http://hbase.apache.org/.} as underlying software stacks for coordinating parallel execution processes to build an RSP engine on the cloud computing infrastructure, called CQELS Cloud~\cite{Danh:2013}. To tailor the RDF-based data processing operations on edge devices (e.g, ARM CPU, Flash-storage), CQLES can be integrated in  RDF4Led~\cite{Anh:2018}, a RISC style RDF engine for lightweight edge devices, to build Fed4Edge~\cite{Manh:2019}. The whole Fed4Edge is smaller than 10MB and needs only 4--6 MB of RAM to process millions of triples on various small devices such as BeagleBone,\footnote{https://beagleboard.org/bone} Raspberry PI.\footnote{https://www.raspberrypi.org/}
Therefore, \system includes an adaptive federation mechanism to enable the coordination of different hardware resources to process query processing pipelines by cooperatively delegating partial workloads to their peer agents.

The Adaptive Federator acts as the query rewriter, which adaptively divides
the input query into sub queries. The rewriter then pushes down the operators as close to the streaming nodes as possible by following the predicate pushdown practice 
%\kien{this is hard to read what it is} 
in common logical optimisation algorithms. 
The metadata subscribed by the other CQELS instances is stored locally.
Similar to~\cite{Dell:2017}, such metadata allows the endpoint services of a CQELS engine to be discovered via the Adaptive Federator. When the Adaptive Federator sends out a subquery, it notifies the Stream Input Handler to subscribe and listens to the results returning from the subquery. On the other hand, the Stream Output
Handler sends out the subqueries to other nodes or sends back the results to the
requester.

Similar to cloud integration with CQELS Cloud~\cite{Danh:2013}, this federation design also covers elastic-scale delopment by using the new development 
of Apache Flink under BIFOLD project. In particular, we use 
the EMMA compiling and parallelizing for data flow systems in ~\cite{Gabor:2021} to scale and optimize our processing pipelines in the cloud infrastructure. 
This will lay the foundation for an integration of the adaptive optimizer with 
the cloud-based stream scheduler and operation allocations.

With the support of the above subscription and discovery operations, a stream
processing pipeline written in CQELS-QL can be deployed across several sites
distributed in different locations: e.g., weather stations provide environmental
sensory streams in various locations on earth. Each autonomous CQELS node
gives access to data streams fed from the streaming nodes connecting to it. Such
stream nodes can ingest a range of sensors, such as air temperature, humidity
and carbon monoxide. When the stream data arrives, this CQELS node can
partially process the data at its processing site and then forward the results as mapping or RDF stream elements to its parent node.

\bibliographystyle{spmpsci}
\bibliography{main.bib}
\end{document}